\documentclass[%
 prl, reprint,
 amsmath,amssymb,
 aps,
 ]{revtex4-2}
\usepackage{graphicx}
\usepackage{comment}
\usepackage[svgnames]{xcolor}
\usepackage{dcolumn}
\usepackage{bm}
\usepackage{verbatim}
\usepackage{commath} 
\usepackage{lpic}
\usepackage{float}
\usepackage{natbib}

\newcommand{\Z}{\mathbb{Z}} 

\begin{document}

\preprint{APS/123-QED}

\title{Observability of a Sharp Majorana Transition in a Few-Body Model}


\author{Jared E. Bland}
\email{bland3@purdue.edu}
\affiliation{Department of Physics \& Astronomy, Purdue University, West Lafayette, IN 47907, USA } 
 
 \author{Chris H. Greene}
 \email{chgreene@purdue.edu}
 \affiliation{Department of Physics \& Astronomy, Purdue University, West Lafayette, IN 47907, USA } 
 \affiliation{Purdue Quantum Science and Engineering Institute,Purdue University, West Lafayette, Indiana 47907, USA}
 \author{Birgit Wehefritz-–Kaufmann}%
 \email{ebkaufma@purdue.edu}
\affiliation{Department of Mathematics, Purdue University, West Lafayette, IN 47907, USA}
\affiliation{Department of Physics \& Astronomy, Purdue University, West Lafayette, IN 47907, USA } 
\affiliation{Purdue Quantum Science and Engineering Institute,Purdue University, West Lafayette, Indiana 47907, USA}

\date{\today}

\begin{abstract}

We propose experimentally observable signatures of of topological Majorana quasiparticles in the few-body limit of the interacting cold-atom model of [Iemini et al. Phys. Rev. Lett. 118 200404 (2017)].  In this limit, the total on-site density and single-body correlations change smoothly with the model parameters, while the calculated mutual information of opposite ends of the lattice indicates a sharp transition of the system to a topological ground state.  Furthermore, local density and parity measurements provide an experimentally viable path for observing the ground state Majorana quasiparticles in ultracold atoms.  Our results lay out a promising future for utilizing few-body systems as a testing ground for Majorana physics. 


\end{abstract}


\maketitle



The cold atom community has made remarkable strides in engineering quantum systems to exhibit many body phenomena that were formerly limited to condensed matter systems.  Particularly for lattice systems exhibiting topological many-body phenomena, recent advances in engineered Hamiltonia for cold atoms have enabled large degrees of control over many-body systems for studying these topological phases of matter \cite{TopBandsUCAtoms}. This degree of control of experimental systems extends to the ability to model a particular Hamiltonian and control the interactions of its constituents (see \cite{Schfer2020} and the hundreds of references therein).

Within the many-body community, there has been a wide-spread search for Majorana edge states, to find fermionic quasiparticles whose fermionic creation or annihilation operators exist as sums of two spatially separated Majorana operators, each commuting with the Hamiltonian \footnote{ More specifically, these are the strong zero modes described in \cite{Fendley} and studied in \cite{Maceira_2018,Kemp_2017,Monthus_2018} to name just a few references.}. 
In Kitaev's landmark paper \cite{KitaevWire}, he demonstrated that a simple mean-field model of superconductivity can exhibit these edge-mode Majoranas as quasiparticles in a quantum wire with zero excitation energy in the thermodynamic limit. For finite lattices, an energy gap proportional to $e^{-L}$, where $L$ is the lattice length, is expected for spatially separated Majorana quasiparticles. While there is no short supply of theoretical models of number-conserving systems that support Majorana edge modes \cite{ZollerMQP, Zhou, IeminiDoubleWire, LangBucher, GutherLangBucher, ZollerDoubleWire, ZollerMajColdAtomWire, PhysRevB.84.094503, PhysRevB.84.195436, PhysRevLett.113.267002, PhysRevLett.120.156802, PhysRevLett.106.220402, PhysRevB.84.144509, PhysRevLett.114.100401, PhysRevB.96.085133, Nori, TopSF}, whether synthetic degrees of freedom truly allow such quasiparticles to exist in a quasi-2d lattice have yet to be seen in experiments. 

Interactions play a crucial role in determining the topological phase of a system of fermions \cite{FidkowskiKitaev}.  In recent years, interacting double wire models have been quite fruitful for exploring number-conserving Majorana systems \cite{IeminiDoubleWire, LangBucher, GutherLangBucher, ZollerDoubleWire, ZollerMajColdAtomWire, PhysRevB.84.094503, PhysRevB.84.195436}. In \cite{ZollerMQP}, the double-wire models of \cite{LangBucher, GutherLangBucher, IeminiDoubleWire} were translated into a realistic cold-atom setting with neutral atoms. At roughly the same time, Zhou \textit{et al}. proposed nearly the same model and studied the phase diagram in the many-body limit using a complementary set of parameters \cite{Zhou}.  Several of the signatures seen in \cite{ZollerMQP} survive in the few-body limit and can be shown to emerge as the two-body interaction strength via spin-exchange is increased relative to the optical lattice tunneling rate, allowing the study of the topological phase transition into a ground state with Majorana quasiparticles. 

In this Letter, we use similar parameter values as \cite{ZollerMQP} to verify the few-body limit and examine additional properties of the model, including signatures of the topological phase using quantum mutual information to indicate non-local physics on opposite ends of a 1-d chain. Guided by the transition seen in the mutual information, we then propose experimental signatures to contrast the topologically trivial/non-trivial phases and calculate timescales to study the transition. We conclude with an investigation of the consistency of the ground state with finite-energy Majorana quasiparticles that are gapped due to the finite size effects.


The model of Iemini et al. has a basis that is composed of four hyperfine states of a fermionic alkaline earth-like atom \cite{TwoElectronAtoms}, physically labeled according to an orbital angular momentum quantum number (labeled $p=\pm)$ and pseudo-spin (labeled $\alpha = \uparrow, \downarrow$), with couplings between two effective nuclear spin states and two orbital spin states provided by interactions with the photon field of the optical lattice as well as two-body interactions between the atoms \cite{GaugeUCAtoms, ControllingInteraction, QMMatterSynth}. The Iemini Hamiltonian is constructed from a sum of a nearest-neighbor tunneling, onsite potentials, synthetic spin-orbit coupling, and a two-body spin-exchange interaction:  
\begin{align*}
H_{I}=\sum_{j} H_{T,j} + H_{U,j} + H_{SO,j} + H_{W,j} .    
\end{align*}
The single site tunneling terms are given by 
\begin{align*}
H_{T,j} = \sum_{p, \alpha}  T(a_{\alpha,p,j+1}^\dagger a_{\alpha,p,j} + \mathcal{H.C.}).   
\end{align*}
The local on-site diagonal interaction is composed of terms
\begin{align*}
   H_{U,j} = & U_{+}n_{\uparrow,+1,j}n_{\downarrow, +,j} + U_{-}n_{\downarrow, -, j} n_{\uparrow, -, j} \\
  + & U\sum_{\alpha,\beta}( n_{\alpha,-, j}n_{\beta,+,j}).   
\end{align*}
In recent years, spin-orbit coupling has become feasible using synthetic magnetic fields \cite{WallSynthSOcoupling, SOCferms, SynthDimClockTrans}. The spin-orbit terms may be written as
\begin{align*}
    H_{SO,j} & = (b+\alpha_R)(a_{\uparrow, +, j}^\dagger a_{\downarrow, -, j+1} + a_{\uparrow, -, j}^\dagger a_{\downarrow, +,j+1} + \mathcal{H.C.} ) \\
    &  + (b-\alpha_R)(a_{\uparrow, +, j+1}^\dagger a_{\downarrow, -,j} + a_{\uparrow, -, j+1}^\dagger a_{\downarrow, +, j}  + \mathcal{H.C.} ),
\end{align*}
where the time-reversal symmetry breaking coefficients $b \pm \alpha_R$ describe the sum and difference of an effective Zeeman-splitting term $b$ that is spatially uniform in this approximation and a Rashba velocity $\alpha_R$, providing an analogous effect to an external magnetic field. The final term, a two-body scattering term related to spin-exchange \cite{LocalMagMom, ObsSpinEx}, is given by 
\begin{align*}
    H_{W,j} =  W( a_{\uparrow, +,j}^\dagger a_{\downarrow, -, j}^\dagger {a_{\downarrow, +, j}} {a_{\uparrow, -, j}}  + \mathcal{H.C.}  ).
\end{align*}

With these interactions, the Hamiltonian may be written in a two-block diagonal form according to a suitable parity operator. We propose a slightly different parity operator from the original paper \cite{ZollerMQP}, whose definition needs modification for odd total particle number. In analogy with \cite{LangBucher}, a study of a double-wire model, fermions of species $(\uparrow, +)$ and $(\downarrow, -)$ at site $j$ are analogous to adjacent-site single species fermions in the upper wire of the double-wire model.  With this, we may define two reasonable parity measurements, $P_+$ and $P_-$:
\begin{align}
P_+ = &  \Bigg(\sum_j n_{\uparrow,+,j} + n_{\downarrow,-,j}\Bigg)\bmod{2}, \text{ and} \\
P_- = & \Bigg(\sum_j n_{\uparrow,-,j} + n_{\downarrow,+,j}\Bigg)\bmod{2}.   
\end{align}
Importantly, these two definitions are equivalent for even total particle number, but interchange the meaning of even- and odd-parity of \cite{ZollerMQP}.  For odd lattice fillings, $P_+$ and $P_-$ define even- and odd-parities in the opposite manner. For this paper, we use the definition $P_+$. 

The Hamiltonian has a chiral symmetry that leaves the Hamiltonian invariant: $a_{\alpha,p,j}^{(\dagger)} \mapsto a_{-\alpha, p, L-j}^{(\dagger)}$. For even numbered fillings, this symmetry commutes with the degenerate parity operators; however, in lattices with odd total occupation number, the chiral transformation exchanges the parity sector of a state for every eigenstate of the parity operator.  Thus, for all parity eigenstates with odd total particle number, the energy spectrum is at least doubly degenerate due to the chiral symmetry relationship between states in opposite parity sectors. For states with even total particle number, the $\Z_2$ symmetry from the generalized notion of parity commutes with the chiral transformation, and the states are not identically degenerate. When $U_{\pm}, U =0$, there is an additional symmetry $a_{\alpha,p,j}^{(\dagger)} \mapsto a_{-\alpha, -p, j}^{(\dagger)}$ that simultaneously commutes with the chiral transformation as well as the parity operator, and so does not place any additional constraints on the spectrum irrespective of the particle number. 


%
%
For the remainder of this paper, we set $T=-1$ and consistently use $W/T$ in figures to emphasize that $T$ determines the energy scale used, and further set $(U_\pm, U,b+\alpha_R, b-\alpha_R)=(0,0,8,0)$ to utilize similar parameters as \cite{ZollerMQP} with $L=7=N$ in the even parity sector. We will study the effect of diagonal interactions ($U,U_{\pm}\neq 0$) and weaker time reversal symmetry breaking ($\alpha_R \neq \pm b$) in future work, but have included those terms for completeness of the model. $W/T$ is varied to study the transition from the non-topological regime with no edge modes to the topological regime with edge-state physics.

In the few-body limit, the spectrum is discretized and the transition regime is spread out over a large set of parameter values as a finite-size effect.  From Fig. \ref{LowESpectrum}, we can see that in the strong-coupling limit of the few-body regime, the ground state manifold approaches being truly degenerate, while the next nearly-degenerate set of energies crosses very slowly with a tuning of $W/T$. 



\begin{figure}[ht]
\begin{lpic}[l(3mm),r(5mm),t(5mm),b(1mm),draft,clean]{ShiftedLowEnergySpectraL7N7(7.5cm)}
    %
    %
    \lbl[l]{0,35,90; $E/T - Q_3/T $}
    \lbl[b]{200,35; $\frac{W}{T}$}
    %
    %
    \lbl[l]{92,82,90; $\frac{E}{T} - \frac{Q_3}{T}$ }
    \lbl[b]{155,77; $\frac{W}{T}$}
    %
    %
    \lbl[l]{20,75;  $\mathbin{\textcolor{Navy}{\square}}$ Gnd. State}
    \lbl[l]{21,85;  $\mathbin{\textcolor{Blue}{\star}}$ $1^\text{st}$ Exc.}
    \lbl[l]{21,95;  $\mathbin{\textcolor{Magenta}{\circ}}$ $2^\text{nd}$ Exc.}
    \lbl[l]{20,105;  $\mathbin{\textcolor{OrangeRed}{+}}$ $3^\text{rd}$ Exc.}
    \lbl[l]{21,115;  \small{$\mathbin{\textcolor{Pink}{\circ}}$} $4^\text{th}$ Exc.}
\end{lpic}
\caption[CaptionLowE]{ 
(Color online) The lowest energy eigenvalues of the even parity sector, shifted so that the center of the low-lying energies is approximately zero by subtracting a cubic fit of the average of the first four energies.\footnote{ A cubic fit is the lowest degree polynomial fit that gives an accurate approximation to the mean low-energy spectrum as a function of $W/T$ when the problem was solved using exact diagonalization. The  fit $Q_3(W/T)$ is given by $0.0016(W/T)^3 - 0.1303(W/T)^2 + 0.8999(W/T)-64.131$. } \\ During the crossover from the non-topological to the topological phase, a zoo of states is formed and we can see the four lowest states separate, with the ground state becoming approximately (doubly) degenerate. We use the same color scheme for all plots and focus on the lowest four energy states as these are the states that interact with the lowest band in the quasi-adiabatic few-body limit. \\ 
\textit{Inset:} There is an avoided crossing between the lowest energy state and the second excited state. When $U=0=U_{\pm}$, the Hamiltonian may be written in a four-block diagonal basis, and there is a non-avoided crossing with the first excited state, which lies in another symmetry sector (but becomes coupled when on-site interacts are included). 
%
%
}
\label{LowESpectrum}
\end{figure}

\begin{figure}[ht]
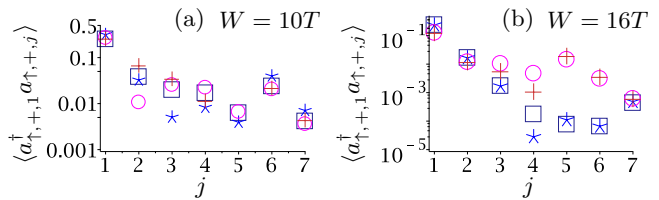

\begin{lpic}[l(3mm),r(0mm),t(3mm),b(2mm)]{L7N7W10GF(39.5mm)} 
    \lbl[l]{90,110; (a)}
    \lbl[l]{120,110;$W=10T$}
    \lbl[l]{-8,15,90;$\langle a_{\uparrow,+,1}^\dagger a_{\uparrow,+,j} \rangle $ }
    \lbl[b]{108,-7;$j$}
\end{lpic}
\begin{lpic}[l(3mm),r(0mm),t(3mm),b(2mm)]{L7N7W16GF(39.5mm)} 
    \lbl[l]{90,110; (b)}
    \lbl[l]{120,110;$W=16T$}
    \lbl[l]{-8,15,90;$\langle a_{\uparrow,+,1}^\dagger a_{\uparrow,+,j} \rangle $}
    \lbl[b]{108,-7;$j$}
\end{lpic}
\caption{Single species correlators $\langle a_{\uparrow,+,1}^\dagger a_{\uparrow,+,j} \rangle $ are shown for the four lowest energy states for $(L,N,T,U,\alpha_R,b)=(7,7,-1,0,4,4)$ (refer to Fig \ref{LowESpectrum} for the legend). (a) In the trivial phase ($W=10T$), all low-energy states exhibit a staggered decay with no rebounding of single-particle correlations. (b) In the topological phase ($W=16T$) the degenerate ground state manifold exhibits exponential decay into the bulk and rebounding on the other end of the chain while the trivial states exhibit correlations that do not rebound. As $W$ is increased in magnitude, the bulk continues to lose correlation with the edges.}
\label{GF}
\end{figure}

Our few-body calculations show strong qualitative agreement with \cite{ZollerMQP}. For $W/T = 16$, the ground state single species correlators $\langle a_{\uparrow, +,1}^\dagger a_{\uparrow, - , j} \rangle$, which measure the intra-species correlation of site 1 with site $j$, initially decay, then rebound exponentially, see Fig. \ref{GF} (b). On the other hand, the non-topological low-energy states decay exponentially with no rebounding, as shown in Fig. \ref{GF} (a) for $W=10T$ as well as the non-topological excited states of Fig. \ref{GF} (b).  The single species Green functions indicate long-range correlations as $W/T$ is increased but do not change abruptly in the few-body limit, making it difficult to use the single-species correlations as the sole signature of the phase transition. 

The edge-edge mutual information is a signature that demonstrates the topological phase transition clearly in the few-body limit. The mutual information of two quantum systems $A,C$ is the relative von Neumann entropy between their joint reduced density matrix and the tensor product of their individual reduced density matrices: $I(A:C) = S_{A}+ S_{C} - S_{AC}$, where $S_{X}$ is the von Neumann entropy of the state's reduced density matrix over region $X$ \cite{Wilde13, NielsenChueng12}.  In the case of a pure state in a quantum system across regions $A,B,C$, we may write this as $I(A\colon C)= S_{A} + S_{C} - S_{B}$ \cite{QIcontvar}.  Fig. \ref{L7MILA1} contains a sample 7 site lattice divided into regions $A,B,C$ for various sizes of the regions, with $L_A=2=L_C$ given in the drawing's annotation.  The mutual information displays a sharp transition as $W/T$ is increased when subsystems $A$ and $C$ comprise 1 site each. For larger regions of $A,C$, the increase of $W/T$ does not cause as sharp of a change in the mutual information between subsystems $A$ and $C$ since the ``bulk'' (region $B$) is providing most of the entanglement across the division via short-ranged entanglement, which varies smoothly with changes in parameters, as shown in Fig. \ref{L7MILA3}. For the case of $N=7$ with no nonzero potential terms, the chiral symmetry guarantees that the entanglement spectra in the center of the lattice are identical in the even and odd parity ground states (similar to \cite{TurnerPollmannBerg}'s inversion symmetry analysis of a mean-field model). Thus, the mutual information gives insight into the interplay between the length scales of edge-modes and short-ranged bulk entanglement even in the presence of additional symmetries that prevent the analysis using the entanglement spectra in opposite parity sectors. 
\begin{figure}[ht]
\textbf{7 Site Lattice with Several Sample Divisions}\par\medskip
\includegraphics[width=7.5 cm] {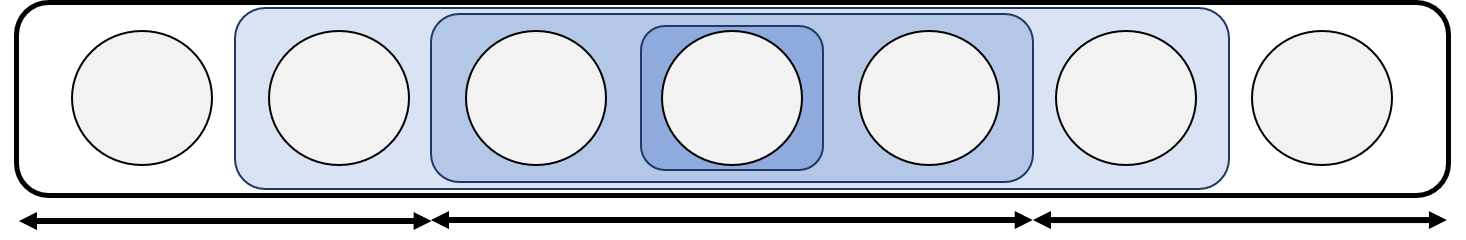} 

\begin{lpic}[l(5mm),r(5mm),t(5mm),b(2mm)]{L7N7MILA1(75mm)} 
    \lbl[l]{-8,50,90;$I(1\colon 7)$}
    \lbl[b]{105,-2;$W/T$}
    
    \lbl[b]{30,122;$L_A=2$}
    \lbl[b]{95,122;$L_B=3$}
    \lbl[b]{165,122;$L_C=2$}
\end{lpic}
\caption{\textit{Top}: Sample lattice with division into three regions $A$, $B$, and $C$. \textit{Bottom}: Mutual information between the first and last sites ($L_A$=$1$=$L_C$) for the four low-energy states as $W/T$ is varied (refer to Fig \ref{LowESpectrum} for the legend). The degenerate ground state manifold has topological properties after the transition region $W/T \in [11,14]$, while the first excited state manifold has no edge-edge correlation beyond a product state.}
\label{L7MILA1}
\end{figure}
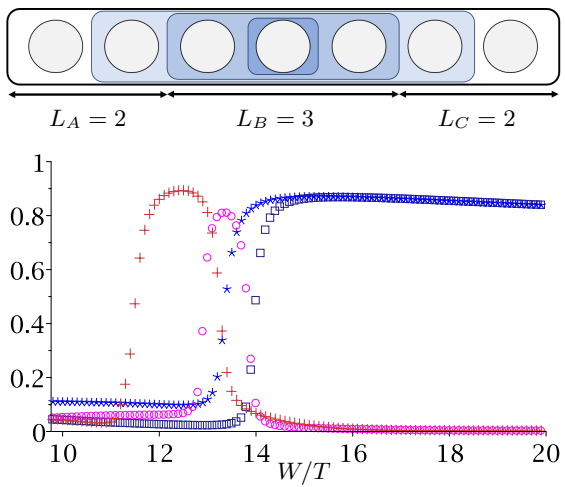
\begin{figure}[ht]
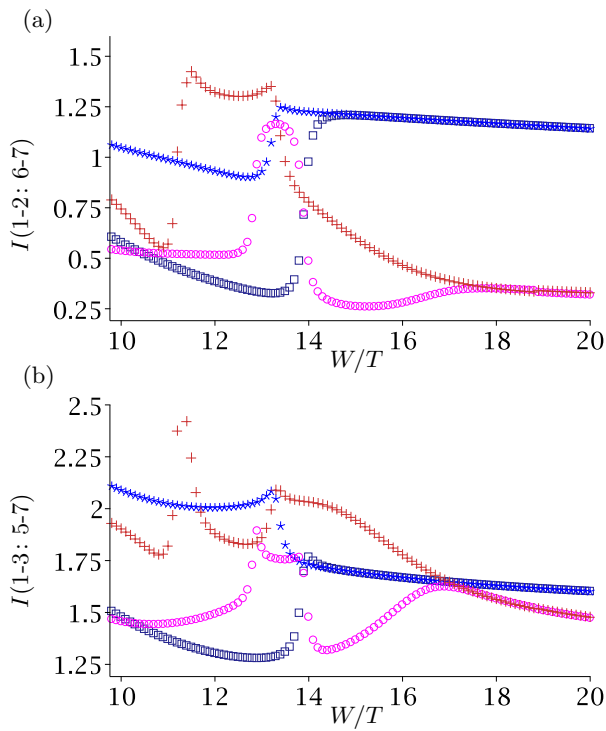

\begin{lpic}[l(5mm),r(5mm),t(3mm),b(1mm)]{L7N7MILA2(75mm)} 
    \lbl[l]{-8,40,90;$I(1$-$2 \colon 6$-$7)$}
    \lbl[b]{108,-2;$W/T$}
    \lbl[l]{-8,120; (a)}
\end{lpic}

\begin{lpic}[l(5mm),r(5mm),t(1mm),b(2mm)]{L7N7MILA3(75mm)} 
    \lbl[l]{-8,40,90;$I(1$-$3 \colon 5$-$7)$}
    \lbl[b]{108,-2;$W/T$}
    \lbl[l]{-8,120; (b)}
\end{lpic}
\caption{Mutual information of the left and right edges for (a) $L_A$=$2$=$L_C$ and (b) $L_A$=$3$=$L_C$. As the widths of the edges for the mutual information calculations are increased, 
one can see the evolution towards bulk behavior as short-ranged entanglement provides the majority of the entanglement between the two halves with finite-size effects causing the cusp-like features in the transition region. }
\label{L7MILA3}
\end{figure}

While the mutual information provides a theoretical understanding of the non-local nature of the ground state manifold in the topological regime, it is not experimentally measurable since it is nonlinear in the density matrix. Equally important, although it provides a measure of topological physics by integrating out the ``bulk'' center region of the lattice, the edge-edge mutual information does not give direct insight into any quasiparticle properties of the ground state aside from non-local edge-edge entanglement. 

In addition to the single particle Green functions of Fig. \ref{GF} (a)-(b),  we propose two additional  experimental signatures distinguishing the topological and non-topological regimes of the model: the density expectation values across the lattice and local parity measurements.  

\begin{figure}[h]
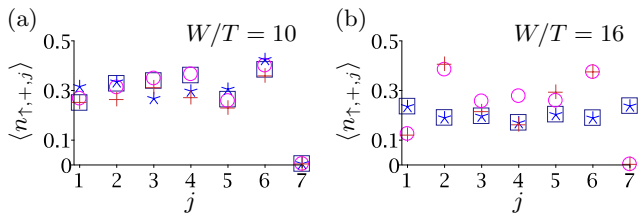

\begin{lpic}[l(3mm),r(0mm),t(3mm),b(2mm)]{L7N7W10Density(39.5mm)} 
    \lbl[l]{-18,120; (a)}
    \lbl[l]{-9,40,90;$\langle n_{\uparrow,+,j} \rangle$}
    \lbl[l]{100,110;$W/T=10$}
    \lbl[b]{103,-7;$j$}
\end{lpic}
\begin{lpic}[l(3mm),r(0mm),t(3mm),b(2mm)]{L7N7W16Density(39.5mm)} 
    \lbl[l]{-18,120; (b)}
    \lbl[l]{-9,40,90;$\langle n_{\uparrow,+,j} \rangle$}
    \lbl[l]{100,110;$W/T=16$}
    \lbl[b]{103,-7;$j$}
\end{lpic}
\caption{(a) In the trivial regime of the model ($W=10T$ in this plot), all low-energy states have very low densities at one edge of the lattice for each particle species (refer to Fig \ref{LowESpectrum} for the legend). (b) In the topological regime ($W=16T$), the densities of the topological ground state manifold have a comparable values on both ends of the lattice, while the non-topological excited states maintain near-zero edge densities at one end of the lattice.} 
\label{Density}
\end{figure}

The density expectation values $\langle n_{\alpha, \pm, j} \rangle $ distinguish between topological and non-topological states when varying $W/T$. Below the transition threshold (indicated by the change in the  mutual information), all four fermionic species have nearly 0 density at one end of the lattice or the other for all low-energy eigenvectors (though only $\langle n_{\uparrow, +, j} \rangle$ is plotted here), exemplified in Fig. \ref{Density} (a) with $W/T=10$. When the mutual information indicates a topological ground state, the densities at opposite ends have similar magnitudes, with the densities of the ground state manifold at $W/T=16$ in Fig. \ref{Density} (b) being characteristic examples. 

\begin{figure}[h]
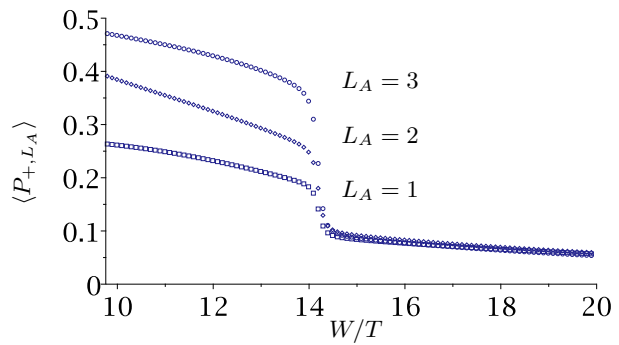

\begin{lpic}[l(5mm),r(5mm),t(0mm),b(1mm),draft,clean]{L7N7ParExpectComparison(75mm)} 
    \lbl[l]{-10,45,90;$\langle P_{+,L_A} \rangle$}
    \lbl[l]{100,88;$L_A=3$}
    \lbl[l]{100,69;$L_A=2$}
    \lbl[l]{100,50;$L_A=1$}
    \lbl[b]{105,-3;$W/T$}
\end{lpic}
\caption{Local ground state parity measurements of the left-hand side $\langle P_{+,L_A} \rangle$ plotted for $L_A=1,2,3$ in the even parity sector.  With just a single site, the local parity measure shows a small but noticeable transition, while the local parity of the first three sites provides a very strong evidence of the phase transition.}
\label{ParExpectComp}
\end{figure}

A sharp few-body experimental signature of the proposed Majorana mode is provided by the parity operator restricted to the first $L_A$ sites on the left-hand side of the lattice: $P_{+,L_A}= (\sum_{j=1\dots L_A} n_{\uparrow, +,j} + n_{\downarrow, -, j})\bmod{2}$. Since two spatially separated Majorana modes each locally break parity, but conserve it globally, we may look at the expectation value of the operator $P_{+,L_A}$ while keeping $L_A < L/2$. With only 7 sites, this signature provides evidence of edge parity breaking while maintaining global parity. In the even parity sector, the parity expectation on the left- and right-hand sides of any cut are identical (to cancel and give 0 as the global parity measurement). The even parity ground state sharply changes in its parity expectation value for the left-hand side, as shown in Fig \ref{ParExpectComp}, which plots $\langle P_{+,L_A} \rangle$ of the even parity ground state for $L_A=1,2,3$. This sharp transition of the left-hand side parity measurement while fixing the global parity gives a sharp experimental signature of a topological mode strongly consistent with Majorana quasiparticles in the ground state.

The presence of an avoided crossing of the energy levels in Fig. \ref{LowESpectrum} as a function of the parameter $W/T$ in the Hamiltonian suggests a way to experimentally create the topological phase possessing Majorana character, starting from the non-topological phase.  Specifically, begin by preparing the system in the lowest energy level to the left of the avoided crossing, i.e. at $W/T \ll 13.9$.  Then by sweeping at a sufficiently slow ramp to $W/T \gg 13.9$, the system will transition into the topological Majorana phase with a controllable high probability.  The requisite ramp speed to stay in the lowest energy level through this avoided crossing can be estimated from Fig. \ref{LowESpectrum} using the Landau-Zener model \cite{Nakamura}. For definiteness, we consider a scenario where the tunneling term is fixed at $T=100$ $h$ Hz, while $W$ is swept through the value $13.9T$ at a linear temporal ramp rate $dW/dt$.  Then the Landau-Zener formula gives the probability to remain in the lower branch of the avoided crossing to be ${\cal P}=1-\exp({-2\pi \Gamma})$, where $\Gamma=\Delta^2 /\hbar/(4|dE_2/dW-dE_1/dW|dW/dt)$.  Our calculation of the critical value of the ramp rate for which the adiabatic transition occurs with 50$\%$ probability is $(dW/dt)_{\rm crit} = 1360\  h {\rm Hz/s}$, i.e. $\Gamma=(0.110\times 1360\  h {\rm Hz/sec})/dW/dt$.  Thus a ramp speed of $136\ h {\rm Hz/s}$ keeps the system on the lower branch and produces the topological Majorana phase with $99.9\%$ probability. For Yb, the lifetime of the excited electronic states ($\sim 10^2$ sec) is far longer than the time to sweep through a change of $W$ of $1000\ h {\rm Hz}$ with the slow ramping rate of $136\ h {\rm Hz/s}$ \cite{WallSynthSOcoupling, SOCferms}.  The long lifetime along with the Landau-Zener analysis gives experimentalists a wide range of control over the transition from the non-topological phase to the topological Majorana phase.


We have discussed the symmetries and corresponding operators of the interacting model of \cite{ZollerMQP} and proposed a different parity definition in analogy with \cite{LangBucher}.  Furthermore, we have demonstrated the utility of mutual information as a way to complement entanglement spectra for categorizing topological phase in the presence of additional symmetries.  The mutual information together with the local parity measurement in the ground state supports the conclusion of \cite{ZollerMQP} that as the two-body scattering is increased, edge modes consistent with Majorana quasiparticles appear in the ground state.

Future work will yield more insight into the characteristics of the edge state physics. Theoretically, further study of the transition region will lead to estimates of the length scales of the Majorana quasiparticles in the ground state 
as well as stability under one-body on-site noise. The experimental signatures proposed here focus on local density measurements, which are standard in 
ultracold atomic experimental groups, and are accessible to groups studying these alkaline earth-like atoms \cite{SynthDimClockTrans, WallSynthSOcoupling, SOCferms, DirectObsChiral, LocalMagMom, ObsSpinEx, SchemeSingleAtom, YbClockTrans}.  In the few-body limit of only 7 sites and 7 particles, our results suggest that there are non-topological excited states gapped away from the degenerate Majorana-like topological ground state manifold due to finite-size effects.


We would like to thank Nima Lashkari for many helpful discussions on quantum information theory.  This work has been partially supported by NSF PHY-1255409 (J.E.B. and B.W.-K.) and NSF PHY-1912350 (C.H.G.).



\providecommand{\noopsort}[1]{}\providecommand{\singleletter}[1]{#1}%
\begin{thebibliography}{43}%
\makeatletter
\providecommand \@ifxundefined [1]{%
 \@ifx{#1\undefined}
}%
\providecommand \@ifnum [1]{%
 \ifnum #1\expandafter \@firstoftwo
 \else \expandafter \@secondoftwo
 \fi
}%
\providecommand \@ifx [1]{%
 \ifx #1\expandafter \@firstoftwo
 \else \expandafter \@secondoftwo
 \fi
}%
\providecommand \natexlab [1]{#1}%
\providecommand \enquote  [1]{``#1''}%
\providecommand \bibnamefont  [1]{#1}%
\providecommand \bibfnamefont [1]{#1}%
\providecommand \citenamefont [1]{#1}%
\providecommand \href@noop [0]{\@secondoftwo}%
\providecommand \href [0]{\begingroup \@sanitize@url \@href}%
\providecommand \@href[1]{\@@startlink{#1}\@@href}%
\providecommand \@@href[1]{\endgroup#1\@@endlink}%
\providecommand \@sanitize@url [0]{\catcode `\\12\catcode `\$12\catcode
  `\&12\catcode `\#12\catcode `\^12\catcode `\_12\catcode `\%12\relax}%
\providecommand \@@startlink[1]{}%
\providecommand \@@endlink[0]{}%
\providecommand \url  [0]{\begingroup\@sanitize@url \@url }%
\providecommand \@url [1]{\endgroup\@href {#1}{\urlprefix }}%
\providecommand \urlprefix  [0]{URL }%
\providecommand \Eprint [0]{\href }%
\providecommand \doibase [0]{https://doi.org/}%
\providecommand \selectlanguage [0]{\@gobble}%
\providecommand \bibinfo  [0]{\@secondoftwo}%
\providecommand \bibfield  [0]{\@secondoftwo}%
\providecommand \translation [1]{[#1]}%
\providecommand \BibitemOpen [0]{}%
\providecommand \bibitemStop [0]{}%
\providecommand \bibitemNoStop [0]{.\EOS\space}%
\providecommand \EOS [0]{\spacefactor3000\relax}%
\providecommand \BibitemShut  [1]{\csname bibitem#1\endcsname}%
\let\auto@bib@innerbib\@empty
\bibitem [{\citenamefont {Cooper}\ \emph {et~al.}(2019)\citenamefont {Cooper},
  \citenamefont {Dalibard},\ and\ \citenamefont {Spielman}}]{TopBandsUCAtoms}%
  \BibitemOpen
  \bibfield  {author} {\bibinfo {author} {\bibfnamefont {N.~R.}\ \bibnamefont
  {Cooper}}, \bibinfo {author} {\bibfnamefont {J.}~\bibnamefont {Dalibard}},\
  and\ \bibinfo {author} {\bibfnamefont {I.~B.}\ \bibnamefont {Spielman}},\
  }\bibfield  {title} {\bibinfo {title} {Topological bands for ultracold
  atoms},\ }\href@noop {} {\bibfield  {journal} {\bibinfo  {journal} {Rev. Mod.
  Phys.}\ }\textbf {\bibinfo {volume} {91}},\ \bibinfo {pages} {015005}
  (\bibinfo {year} {2019})}\BibitemShut {NoStop}%
\bibitem [{\citenamefont {Sch\"{a}fer}\ \emph {et~al.}(2020)\citenamefont
  {Sch\"{a}fer}, \citenamefont {Fukuhara}, \citenamefont {Sugawa},
  \citenamefont {Takasu},\ and\ \citenamefont {Takahashi}}]{Schfer2020}%
  \BibitemOpen
  \bibfield  {author} {\bibinfo {author} {\bibfnamefont {F.}~\bibnamefont
  {Sch\"{a}fer}}, \bibinfo {author} {\bibfnamefont {T.}~\bibnamefont
  {Fukuhara}}, \bibinfo {author} {\bibfnamefont {S.}~\bibnamefont {Sugawa}},
  \bibinfo {author} {\bibfnamefont {Y.}~\bibnamefont {Takasu}},\ and\ \bibinfo
  {author} {\bibfnamefont {Y.}~\bibnamefont {Takahashi}},\ }\bibfield  {title}
  {\bibinfo {title} {Tools for quantum simulation with ultracold atoms in
  optical lattices},\ }\href {https://doi.org/10.1038/s42254-020-0195-3}
  {\bibfield  {journal} {\bibinfo  {journal} {Nature Reviews Physics}\ }\textbf
  {\bibinfo {volume} {2}},\ \bibinfo {pages} {411–425} (\bibinfo {year}
  {2020})}\BibitemShut {NoStop}%
\bibitem [{Note1()}]{Note1}%
  \BibitemOpen
  \bibinfo {note} {More specifically, these are the strong zero modes described
  in \cite {Fendley} and studied in \cite {Maceira_2018,Kemp_2017,Monthus_2018}
  to name just a few references.}\BibitemShut {Stop}%
\bibitem [{\citenamefont {Kitaev}(2001)}]{KitaevWire}%
  \BibitemOpen
  \bibfield  {author} {\bibinfo {author} {\bibfnamefont {A.~Y.}\ \bibnamefont
  {Kitaev}},\ }\bibfield  {title} {\bibinfo {title} {Unpaired {M}ajorana
  fermions in quantum wires},\ }\href@noop {} {\bibfield  {journal} {\bibinfo
  {journal} {Pyhsics-Uspekhi}\ }\textbf {\bibinfo {volume} {44}},\ \bibinfo
  {pages} {131} (\bibinfo {year} {2001})}\BibitemShut {NoStop}%
\bibitem [{\citenamefont {Iemini}\ \emph {et~al.}(2017)\citenamefont {Iemini},
  \citenamefont {Mazza}, \citenamefont {Fallani}, \citenamefont {Zoller},
  \citenamefont {Fazio},\ and\ \citenamefont {Dalmonte}}]{ZollerMQP}%
  \BibitemOpen
  \bibfield  {author} {\bibinfo {author} {\bibfnamefont {F.}~\bibnamefont
  {Iemini}}, \bibinfo {author} {\bibfnamefont {L.}~\bibnamefont {Mazza}},
  \bibinfo {author} {\bibfnamefont {L.}~\bibnamefont {Fallani}}, \bibinfo
  {author} {\bibfnamefont {P.}~\bibnamefont {Zoller}}, \bibinfo {author}
  {\bibfnamefont {R.}~\bibnamefont {Fazio}},\ and\ \bibinfo {author}
  {\bibfnamefont {M.}~\bibnamefont {Dalmonte}},\ }\bibfield  {title} {\bibinfo
  {title} {Majorana quasiparticle protected by $\mathbb{Z}_2$ angular momentum
  conservation},\ }\href@noop {} {\bibfield  {journal} {\bibinfo  {journal}
  {Phys. Rev. Lett.}\ }\textbf {\bibinfo {volume} {118}},\ \bibinfo {pages}
  {200404} (\bibinfo {year} {2017})}\BibitemShut {NoStop}%
\bibitem [{\citenamefont {Zhou}\ \emph {et~al.}(2017)\citenamefont {Zhou},
  \citenamefont {Pan}, \citenamefont {Liu}, \citenamefont {Zhang},
  \citenamefont {Yi}, \citenamefont {Chen},\ and\ \citenamefont {Jia}}]{Zhou}%
  \BibitemOpen
  \bibfield  {author} {\bibinfo {author} {\bibfnamefont {X.}~\bibnamefont
  {Zhou}}, \bibinfo {author} {\bibfnamefont {J.-S.}\ \bibnamefont {Pan}},
  \bibinfo {author} {\bibfnamefont {Z.-X.}\ \bibnamefont {Liu}}, \bibinfo
  {author} {\bibfnamefont {W.}~\bibnamefont {Zhang}}, \bibinfo {author}
  {\bibfnamefont {W.}~\bibnamefont {Yi}}, \bibinfo {author} {\bibfnamefont
  {G.}~\bibnamefont {Chen}},\ and\ \bibinfo {author} {\bibfnamefont
  {S.}~\bibnamefont {Jia}},\ }\bibfield  {title} {\bibinfo {title}
  {Symmetry-protected topological states for interacting fermions in
  alkaline-earth-like atoms},\ }\href@noop {} {\bibfield  {journal} {\bibinfo
  {journal} {Phys. Rev. Lett.}\ }\textbf {\bibinfo {volume} {119}},\ \bibinfo
  {pages} {185701} (\bibinfo {year} {2017})}\BibitemShut {NoStop}%
\bibitem [{\citenamefont {Iemini}\ \emph {et~al.}(2015)\citenamefont {Iemini},
  \citenamefont {Mazza}, \citenamefont {Rossini}, \citenamefont {Fazio},\ and\
  \citenamefont {Diehl}}]{IeminiDoubleWire}%
  \BibitemOpen
  \bibfield  {author} {\bibinfo {author} {\bibfnamefont {F.}~\bibnamefont
  {Iemini}}, \bibinfo {author} {\bibfnamefont {L.}~\bibnamefont {Mazza}},
  \bibinfo {author} {\bibfnamefont {D.}~\bibnamefont {Rossini}}, \bibinfo
  {author} {\bibfnamefont {R.}~\bibnamefont {Fazio}},\ and\ \bibinfo {author}
  {\bibfnamefont {S.}~\bibnamefont {Diehl}},\ }\bibfield  {title} {\bibinfo
  {title} {Localized {M}ajorana-like modes in a number-conserving setting: An
  exactly solvable model},\ }\href
  {https://doi.org/10.1103/PhysRevLett.115.156402} {\bibfield  {journal}
  {\bibinfo  {journal} {Phys. Rev. Lett.}\ }\textbf {\bibinfo {volume} {115}},\
  \bibinfo {pages} {156402} (\bibinfo {year} {2015})}\BibitemShut {NoStop}%
\bibitem [{\citenamefont {Lang}\ and\ \citenamefont
  {B\"{u}chler}(2015)}]{LangBucher}%
  \BibitemOpen
  \bibfield  {author} {\bibinfo {author} {\bibfnamefont {N.}~\bibnamefont
  {Lang}}\ and\ \bibinfo {author} {\bibfnamefont {H.~P.}\ \bibnamefont
  {B\"{u}chler}},\ }\bibfield  {title} {\bibinfo {title} {Topological states in
  a microscopic model of interacting fermions},\ }\href@noop {} {\bibfield
  {journal} {\bibinfo  {journal} {Phys. Rev. B}\ }\textbf {\bibinfo {volume}
  {92}},\ \bibinfo {pages} {041118(R)} (\bibinfo {year} {2015})}\BibitemShut
  {NoStop}%
\bibitem [{\citenamefont {Guther}\ \emph {et~al.}(2017)\citenamefont {Guther},
  \citenamefont {Lang},\ and\ \citenamefont {B\"{u}chler}}]{GutherLangBucher}%
  \BibitemOpen
  \bibfield  {author} {\bibinfo {author} {\bibfnamefont {K.}~\bibnamefont
  {Guther}}, \bibinfo {author} {\bibfnamefont {N.}~\bibnamefont {Lang}},\ and\
  \bibinfo {author} {\bibfnamefont {H.~P.}\ \bibnamefont {B\"{u}chler}},\
  }\bibfield  {title} {\bibinfo {title} {Ising anyonic topological phase of
  interacting fermions in one dimensions},\ }\href@noop {} {\bibfield
  {journal} {\bibinfo  {journal} {Phys. Rev. B}\ }\textbf {\bibinfo {volume}
  {96}},\ \bibinfo {pages} {121109(R)} (\bibinfo {year} {2017})}\BibitemShut
  {NoStop}%
\bibitem [{\citenamefont {Kraus}\ \emph {et~al.}(2013)\citenamefont {Kraus},
  \citenamefont {Dalmonte}, \citenamefont {Baranov}, \citenamefont
  {L\"auchli},\ and\ \citenamefont {Zoller}}]{ZollerDoubleWire}%
  \BibitemOpen
  \bibfield  {author} {\bibinfo {author} {\bibfnamefont {C.~V.}\ \bibnamefont
  {Kraus}}, \bibinfo {author} {\bibfnamefont {M.}~\bibnamefont {Dalmonte}},
  \bibinfo {author} {\bibfnamefont {M.~A.}\ \bibnamefont {Baranov}}, \bibinfo
  {author} {\bibfnamefont {A.~M.}\ \bibnamefont {L\"auchli}},\ and\ \bibinfo
  {author} {\bibfnamefont {P.}~\bibnamefont {Zoller}},\ }\bibfield  {title}
  {\bibinfo {title} {Majorana edge states in atomic wires coupled by pair
  hopping},\ }\href {https://doi.org/10.1103/PhysRevLett.111.173004} {\bibfield
   {journal} {\bibinfo  {journal} {Phys. Rev. Lett.}\ }\textbf {\bibinfo
  {volume} {111}},\ \bibinfo {pages} {173004} (\bibinfo {year}
  {2013})}\BibitemShut {NoStop}%
\bibitem [{\citenamefont {Jiang}\ \emph
  {et~al.}(2011{\natexlab{a}})\citenamefont {Jiang}, \citenamefont {Kitagawa},
  \citenamefont {Alicea}, \citenamefont {Akhmerov}, \citenamefont {Pekker},
  \citenamefont {Refael}, \citenamefont {Cirac}, \citenamefont {Demler},
  \citenamefont {Lukin},\ and\ \citenamefont {Zoller}}]{ZollerMajColdAtomWire}%
  \BibitemOpen
  \bibfield  {author} {\bibinfo {author} {\bibfnamefont {L.}~\bibnamefont
  {Jiang}}, \bibinfo {author} {\bibfnamefont {T.}~\bibnamefont {Kitagawa}},
  \bibinfo {author} {\bibfnamefont {J.}~\bibnamefont {Alicea}}, \bibinfo
  {author} {\bibfnamefont {A.~R.}\ \bibnamefont {Akhmerov}}, \bibinfo {author}
  {\bibfnamefont {D.}~\bibnamefont {Pekker}}, \bibinfo {author} {\bibfnamefont
  {G.}~\bibnamefont {Refael}}, \bibinfo {author} {\bibfnamefont {J.~I.}\
  \bibnamefont {Cirac}}, \bibinfo {author} {\bibfnamefont {E.}~\bibnamefont
  {Demler}}, \bibinfo {author} {\bibfnamefont {M.~D.}\ \bibnamefont {Lukin}},\
  and\ \bibinfo {author} {\bibfnamefont {P.}~\bibnamefont {Zoller}},\
  }\bibfield  {title} {\bibinfo {title} {Majorana fermions in equilibrium and
  driven cold atom quantum wires},\ }\href@noop {} {\bibfield  {journal}
  {\bibinfo  {journal} {Phys. Rev. Lett.}\ }\textbf {\bibinfo {volume} {106}},\
  \bibinfo {pages} {220402} (\bibinfo {year} {2011}{\natexlab{a}})}\BibitemShut
  {NoStop}%
\bibitem [{\citenamefont {Cheng}\ and\ \citenamefont
  {Tu}(2011)}]{PhysRevB.84.094503}%
  \BibitemOpen
  \bibfield  {author} {\bibinfo {author} {\bibfnamefont {M.}~\bibnamefont
  {Cheng}}\ and\ \bibinfo {author} {\bibfnamefont {H.-H.}\ \bibnamefont {Tu}},\
  }\bibfield  {title} {\bibinfo {title} {Majorana edge states in interacting
  two-chain ladders of fermions},\ }\href
  {https://doi.org/10.1103/PhysRevB.84.094503} {\bibfield  {journal} {\bibinfo
  {journal} {Phys. Rev. B}\ }\textbf {\bibinfo {volume} {84}},\ \bibinfo
  {pages} {094503} (\bibinfo {year} {2011})}\BibitemShut {NoStop}%
\bibitem [{\citenamefont {Fidkowski}\ \emph {et~al.}(2011)\citenamefont
  {Fidkowski}, \citenamefont {Lutchyn}, \citenamefont {Nayak},\ and\
  \citenamefont {Fisher}}]{PhysRevB.84.195436}%
  \BibitemOpen
  \bibfield  {author} {\bibinfo {author} {\bibfnamefont {L.}~\bibnamefont
  {Fidkowski}}, \bibinfo {author} {\bibfnamefont {R.~M.}\ \bibnamefont
  {Lutchyn}}, \bibinfo {author} {\bibfnamefont {C.}~\bibnamefont {Nayak}},\
  and\ \bibinfo {author} {\bibfnamefont {M.~P.~A.}\ \bibnamefont {Fisher}},\
  }\bibfield  {title} {\bibinfo {title} {Majorana zero modes in one-dimensional
  quantum wires without long-ranged superconducting order},\ }\href
  {https://doi.org/10.1103/PhysRevB.84.195436} {\bibfield  {journal} {\bibinfo
  {journal} {Phys. Rev. B}\ }\textbf {\bibinfo {volume} {84}},\ \bibinfo
  {pages} {195436} (\bibinfo {year} {2011})}\BibitemShut {NoStop}%
\bibitem [{\citenamefont {Ortiz}\ \emph {et~al.}(2014)\citenamefont {Ortiz},
  \citenamefont {Dukelsky}, \citenamefont {Cobanera}, \citenamefont {Esebbag},\
  and\ \citenamefont {Beenakker}}]{PhysRevLett.113.267002}%
  \BibitemOpen
  \bibfield  {author} {\bibinfo {author} {\bibfnamefont {G.}~\bibnamefont
  {Ortiz}}, \bibinfo {author} {\bibfnamefont {J.}~\bibnamefont {Dukelsky}},
  \bibinfo {author} {\bibfnamefont {E.}~\bibnamefont {Cobanera}}, \bibinfo
  {author} {\bibfnamefont {C.}~\bibnamefont {Esebbag}},\ and\ \bibinfo {author}
  {\bibfnamefont {C.}~\bibnamefont {Beenakker}},\ }\bibfield  {title} {\bibinfo
  {title} {Many-body characterization of particle-conserving topological
  superfluids},\ }\href {https://doi.org/10.1103/PhysRevLett.113.267002}
  {\bibfield  {journal} {\bibinfo  {journal} {Phys. Rev. Lett.}\ }\textbf
  {\bibinfo {volume} {113}},\ \bibinfo {pages} {267002} (\bibinfo {year}
  {2014})}\BibitemShut {NoStop}%
\bibitem [{\citenamefont {Zhang}\ and\ \citenamefont
  {Liu}(2018)}]{PhysRevLett.120.156802}%
  \BibitemOpen
  \bibfield  {author} {\bibinfo {author} {\bibfnamefont {R.-X.}\ \bibnamefont
  {Zhang}}\ and\ \bibinfo {author} {\bibfnamefont {C.-X.}\ \bibnamefont
  {Liu}},\ }\bibfield  {title} {\bibinfo {title} {Crystalline
  symmetry-protected majorana mode in number-conserving dirac semimetal
  nanowires},\ }\href {https://doi.org/10.1103/PhysRevLett.120.156802}
  {\bibfield  {journal} {\bibinfo  {journal} {Phys. Rev. Lett.}\ }\textbf
  {\bibinfo {volume} {120}},\ \bibinfo {pages} {156802} (\bibinfo {year}
  {2018})}\BibitemShut {NoStop}%
\bibitem [{\citenamefont {Jiang}\ \emph
  {et~al.}(2011{\natexlab{b}})\citenamefont {Jiang}, \citenamefont {Kitagawa},
  \citenamefont {Alicea}, \citenamefont {Akhmerov}, \citenamefont {Pekker},
  \citenamefont {Refael}, \citenamefont {Cirac}, \citenamefont {Demler},
  \citenamefont {Lukin},\ and\ \citenamefont
  {Zoller}}]{PhysRevLett.106.220402}%
  \BibitemOpen
  \bibfield  {author} {\bibinfo {author} {\bibfnamefont {L.}~\bibnamefont
  {Jiang}}, \bibinfo {author} {\bibfnamefont {T.}~\bibnamefont {Kitagawa}},
  \bibinfo {author} {\bibfnamefont {J.}~\bibnamefont {Alicea}}, \bibinfo
  {author} {\bibfnamefont {A.~R.}\ \bibnamefont {Akhmerov}}, \bibinfo {author}
  {\bibfnamefont {D.}~\bibnamefont {Pekker}}, \bibinfo {author} {\bibfnamefont
  {G.}~\bibnamefont {Refael}}, \bibinfo {author} {\bibfnamefont {J.~I.}\
  \bibnamefont {Cirac}}, \bibinfo {author} {\bibfnamefont {E.}~\bibnamefont
  {Demler}}, \bibinfo {author} {\bibfnamefont {M.~D.}\ \bibnamefont {Lukin}},\
  and\ \bibinfo {author} {\bibfnamefont {P.}~\bibnamefont {Zoller}},\
  }\bibfield  {title} {\bibinfo {title} {Number conserving theory for
  topologically protected degeneracy in one-dimensional fermions},\ }\href@noop
  {} {\bibfield  {journal} {\bibinfo  {journal} {Phys. Rev. Lett.}\ }\textbf
  {\bibinfo {volume} {106}},\ \bibinfo {pages} {220402} (\bibinfo {year}
  {2011}{\natexlab{b}})}\BibitemShut {NoStop}%
\bibitem [{\citenamefont {Sau}\ \emph {et~al.}(2011)\citenamefont {Sau},
  \citenamefont {Halperin}, \citenamefont {Flensberg},\ and\ \citenamefont
  {Das~Sarma}}]{PhysRevB.84.144509}%
  \BibitemOpen
  \bibfield  {author} {\bibinfo {author} {\bibfnamefont {J.~D.}\ \bibnamefont
  {Sau}}, \bibinfo {author} {\bibfnamefont {B.~I.}\ \bibnamefont {Halperin}},
  \bibinfo {author} {\bibfnamefont {K.}~\bibnamefont {Flensberg}},\ and\
  \bibinfo {author} {\bibfnamefont {S.}~\bibnamefont {Das~Sarma}},\ }\bibfield
  {title} {\bibinfo {title} {Number conserving theory for topologically
  protected degeneracy in one-dimensional fermions},\ }\href
  {https://doi.org/10.1103/PhysRevB.84.144509} {\bibfield  {journal} {\bibinfo
  {journal} {Phys. Rev. B}\ }\textbf {\bibinfo {volume} {84}},\ \bibinfo
  {pages} {144509} (\bibinfo {year} {2011})}\BibitemShut {NoStop}%
\bibitem [{\citenamefont {Ruhman}\ \emph {et~al.}(2015)\citenamefont {Ruhman},
  \citenamefont {Berg},\ and\ \citenamefont {Altman}}]{PhysRevLett.114.100401}%
  \BibitemOpen
  \bibfield  {author} {\bibinfo {author} {\bibfnamefont {J.}~\bibnamefont
  {Ruhman}}, \bibinfo {author} {\bibfnamefont {E.}~\bibnamefont {Berg}},\ and\
  \bibinfo {author} {\bibfnamefont {E.}~\bibnamefont {Altman}},\ }\bibfield
  {title} {\bibinfo {title} {Topological states in a one-dimensional fermi gas
  with attractive interaction},\ }\href
  {https://doi.org/10.1103/PhysRevLett.114.100401} {\bibfield  {journal}
  {\bibinfo  {journal} {Phys. Rev. Lett.}\ }\textbf {\bibinfo {volume} {114}},\
  \bibinfo {pages} {100401} (\bibinfo {year} {2015})}\BibitemShut {NoStop}%
\bibitem [{\citenamefont {Ruhman}\ and\ \citenamefont
  {Altman}(2017)}]{PhysRevB.96.085133}%
  \BibitemOpen
  \bibfield  {author} {\bibinfo {author} {\bibfnamefont {J.}~\bibnamefont
  {Ruhman}}\ and\ \bibinfo {author} {\bibfnamefont {E.}~\bibnamefont
  {Altman}},\ }\bibfield  {title} {\bibinfo {title} {Topological degeneracy and
  pairing in a one-dimensional gas of spinless fermions},\ }\href
  {https://doi.org/10.1103/PhysRevB.96.085133} {\bibfield  {journal} {\bibinfo
  {journal} {Phys. Rev. B}\ }\textbf {\bibinfo {volume} {96}},\ \bibinfo
  {pages} {085133} (\bibinfo {year} {2017})}\BibitemShut {NoStop}%
\bibitem [{\citenamefont {Zhang}\ and\ \citenamefont {Nori}(2016)}]{Nori}%
  \BibitemOpen
  \bibfield  {author} {\bibinfo {author} {\bibfnamefont {P.}~\bibnamefont
  {Zhang}}\ and\ \bibinfo {author} {\bibfnamefont {F.}~\bibnamefont {Nori}},\
  }\bibfield  {title} {\bibinfo {title} {Majorana bound states in a disordered
  quantum dot chain},\ }\href@noop {} {\bibfield  {journal} {\bibinfo
  {journal} {New J. Phys.}\ }\textbf {\bibinfo {volume} {18}},\ \bibinfo
  {pages} {043033} (\bibinfo {year} {2016})}\BibitemShut {NoStop}%
\bibitem [{\citenamefont {Ortiz}\ and\ \citenamefont {Cobanera}(2016)}]{TopSF}%
  \BibitemOpen
  \bibfield  {author} {\bibinfo {author} {\bibfnamefont {G.}~\bibnamefont
  {Ortiz}}\ and\ \bibinfo {author} {\bibfnamefont {E.}~\bibnamefont
  {Cobanera}},\ }\bibfield  {title} {\bibinfo {title} {What is a
  particle-conserving topological superfluid? the fate of {M}ajorana modes
  beyond mean-field theory},\ }\href@noop {} {\bibfield  {journal} {\bibinfo
  {journal} {Annals of Physics}\ }\textbf {\bibinfo {volume} {372}},\ \bibinfo
  {pages} {357–374} (\bibinfo {year} {2016})}\BibitemShut {NoStop}%
\bibitem [{\citenamefont {Fidkowski}\ and\ \citenamefont
  {Kitaev}(2010)}]{FidkowskiKitaev}%
  \BibitemOpen
  \bibfield  {author} {\bibinfo {author} {\bibfnamefont {L.}~\bibnamefont
  {Fidkowski}}\ and\ \bibinfo {author} {\bibfnamefont {A.}~\bibnamefont
  {Kitaev}},\ }\bibfield  {title} {\bibinfo {title} {Effects of interactions of
  the topological classification of free fermion systems},\ }\href@noop {}
  {\bibfield  {journal} {\bibinfo  {journal} {Phys. Rev. B}\ }\textbf {\bibinfo
  {volume} {81}},\ \bibinfo {pages} {134509} (\bibinfo {year}
  {2010})}\BibitemShut {NoStop}%
\bibitem [{\citenamefont {He}\ \emph {et~al.}(2019)\citenamefont {He},
  \citenamefont {Hajiyev}, \citenamefont {Ren}, \citenamefont {Song},\ and\
  \citenamefont {Jo}}]{TwoElectronAtoms}%
  \BibitemOpen
  \bibfield  {author} {\bibinfo {author} {\bibfnamefont {C.}~\bibnamefont
  {He}}, \bibinfo {author} {\bibfnamefont {E.}~\bibnamefont {Hajiyev}},
  \bibinfo {author} {\bibfnamefont {Z.}~\bibnamefont {Ren}}, \bibinfo {author}
  {\bibfnamefont {B.}~\bibnamefont {Song}},\ and\ \bibinfo {author}
  {\bibfnamefont {G.}~\bibnamefont {Jo}},\ }\bibfield  {title} {\bibinfo
  {title} {Recent progresses of ultracold two-electron atoms},\ }\href@noop {}
  {\bibfield  {journal} {\bibinfo  {journal} {J. Phys. B: At. Mol. Opt. Phys.}\
  }\textbf {\bibinfo {volume} {52}},\ \bibinfo {pages} {102001} (\bibinfo
  {year} {2019})}\BibitemShut {NoStop}%
\bibitem [{\citenamefont {Goldman}\ \emph {et~al.}(2014)\citenamefont
  {Goldman}, \citenamefont {J\={u}zeliunas}, \citenamefont {\"{O}hberg},\ and\
  \citenamefont {Spielman}}]{GaugeUCAtoms}%
  \BibitemOpen
  \bibfield  {author} {\bibinfo {author} {\bibfnamefont {N.}~\bibnamefont
  {Goldman}}, \bibinfo {author} {\bibfnamefont {G.}~\bibnamefont
  {J\={u}zeliunas}}, \bibinfo {author} {\bibfnamefont {P.}~\bibnamefont
  {\"{O}hberg}},\ and\ \bibinfo {author} {\bibfnamefont {I.~B.}\ \bibnamefont
  {Spielman}},\ }\bibfield  {title} {\bibinfo {title} {Light-induced gauge
  fields for ultracold atoms},\ }\href@noop {} {\bibfield  {journal} {\bibinfo
  {journal} {Rep. Prog. Phys.}\ }\textbf {\bibinfo {volume} {77}},\ \bibinfo
  {pages} {126401} (\bibinfo {year} {2014})}\BibitemShut {NoStop}%
\bibitem [{\citenamefont {Zhang}\ \emph {et~al.}(2020)\citenamefont {Zhang},
  \citenamefont {Cheng}, \citenamefont {Zhang},\ and\ \citenamefont
  {Zhai}}]{ControllingInteraction}%
  \BibitemOpen
  \bibfield  {author} {\bibinfo {author} {\bibfnamefont {R.}~\bibnamefont
  {Zhang}}, \bibinfo {author} {\bibfnamefont {Y.}~\bibnamefont {Cheng}},
  \bibinfo {author} {\bibfnamefont {P.}~\bibnamefont {Zhang}},\ and\ \bibinfo
  {author} {\bibfnamefont {H.}~\bibnamefont {Zhai}},\ }\bibfield  {title}
  {\bibinfo {title} {Controlling the interaction of ultracold alkaline-earth
  atoms},\ }\href@noop {} {\bibfield  {journal} {\bibinfo  {journal} {Nat. Rev.
  Phys.}\ }\textbf {\bibinfo {volume} {2}},\ \bibinfo {pages} {213–220}
  (\bibinfo {year} {2020})}\BibitemShut {NoStop}%
\bibitem [{\citenamefont {Zhang}\ and\ \citenamefont
  {Zhou}(2017)}]{QMMatterSynth}%
  \BibitemOpen
  \bibfield  {author} {\bibinfo {author} {\bibfnamefont {S.-L.}\ \bibnamefont
  {Zhang}}\ and\ \bibinfo {author} {\bibfnamefont {Q.}~\bibnamefont {Zhou}},\
  }\bibfield  {title} {\bibinfo {title} {Manipulating novel quantum phenomena
  using synthetic gauge fields},\ }\href@noop {} {\bibfield  {journal}
  {\bibinfo  {journal} {J. Phys. B: At. Mol. Opt. Phys.}\ }\textbf {\bibinfo
  {volume} {50}},\ \bibinfo {pages} {222001} (\bibinfo {year}
  {2017})}\BibitemShut {NoStop}%
\bibitem [{\citenamefont {Wall}\ \emph {et~al.}(2016)\citenamefont {Wall},
  \citenamefont {Koller}, \citenamefont {Li}, \citenamefont {Zhang},
  \citenamefont {Cooper}, \citenamefont {Ye},\ and\ \citenamefont
  {Rey}}]{WallSynthSOcoupling}%
  \BibitemOpen
  \bibfield  {author} {\bibinfo {author} {\bibfnamefont {M.~L.}\ \bibnamefont
  {Wall}}, \bibinfo {author} {\bibfnamefont {A.~P.}\ \bibnamefont {Koller}},
  \bibinfo {author} {\bibfnamefont {S.}~\bibnamefont {Li}}, \bibinfo {author}
  {\bibfnamefont {X.}~\bibnamefont {Zhang}}, \bibinfo {author} {\bibfnamefont
  {N.~R.}\ \bibnamefont {Cooper}}, \bibinfo {author} {\bibfnamefont
  {J.}~\bibnamefont {Ye}},\ and\ \bibinfo {author} {\bibfnamefont {A.~M.}\
  \bibnamefont {Rey}},\ }\bibfield  {title} {\bibinfo {title} {Synthetic
  spin-orbit coupling in an optical lattice clock},\ }\href@noop {} {\bibfield
  {journal} {\bibinfo  {journal} {Phys. Rev. Lett.}\ }\textbf {\bibinfo
  {volume} {116}},\ \bibinfo {pages} {035301} (\bibinfo {year}
  {2016})}\BibitemShut {NoStop}%
\bibitem [{\citenamefont {Kolkowitz}\ \emph {et~al.}(2017)\citenamefont
  {Kolkowitz}, \citenamefont {Bromley}, \citenamefont {Bothwell}, \citenamefont
  {Wall}, \citenamefont {Marti}, \citenamefont {Koller}, \citenamefont {Zhang},
  \citenamefont {Rey},\ and\ \citenamefont {Ye}}]{SOCferms}%
  \BibitemOpen
  \bibfield  {author} {\bibinfo {author} {\bibfnamefont {S.}~\bibnamefont
  {Kolkowitz}}, \bibinfo {author} {\bibfnamefont {S.~L.}\ \bibnamefont
  {Bromley}}, \bibinfo {author} {\bibfnamefont {T.}~\bibnamefont {Bothwell}},
  \bibinfo {author} {\bibfnamefont {M.~L.}\ \bibnamefont {Wall}}, \bibinfo
  {author} {\bibfnamefont {G.~E.}\ \bibnamefont {Marti}}, \bibinfo {author}
  {\bibfnamefont {A.~P.}\ \bibnamefont {Koller}}, \bibinfo {author}
  {\bibfnamefont {X.}~\bibnamefont {Zhang}}, \bibinfo {author} {\bibfnamefont
  {A.~M.}\ \bibnamefont {Rey}},\ and\ \bibinfo {author} {\bibfnamefont
  {J.}~\bibnamefont {Ye}},\ }\bibfield  {title} {\bibinfo {title} {Spin-orbit
  coupled fermions in an optical lattice clock},\ }\href@noop {} {\bibfield
  {journal} {\bibinfo  {journal} {Nature}\ }\textbf {\bibinfo {volume} {542}},\
  \bibinfo {pages} {66} (\bibinfo {year} {2017})}\BibitemShut {NoStop}%
\bibitem [{\citenamefont {Livi}\ \emph {et~al.}(2016)\citenamefont {Livi},
  \citenamefont {Cappellini}, \citenamefont {Diem}, \citenamefont {Franchi},
  \citenamefont {Clivati}, \citenamefont {Frittelli}, \citenamefont {Levi},
  \citenamefont {Calonico}, \citenamefont {Catani}, \citenamefont {Inguscio},\
  and\ \citenamefont {Fallani}}]{SynthDimClockTrans}%
  \BibitemOpen
  \bibfield  {author} {\bibinfo {author} {\bibfnamefont {L.~F.}\ \bibnamefont
  {Livi}}, \bibinfo {author} {\bibfnamefont {G.}~\bibnamefont {Cappellini}},
  \bibinfo {author} {\bibfnamefont {M.}~\bibnamefont {Diem}}, \bibinfo {author}
  {\bibfnamefont {L.}~\bibnamefont {Franchi}}, \bibinfo {author} {\bibfnamefont
  {C.}~\bibnamefont {Clivati}}, \bibinfo {author} {\bibfnamefont
  {M.}~\bibnamefont {Frittelli}}, \bibinfo {author} {\bibfnamefont
  {F.}~\bibnamefont {Levi}}, \bibinfo {author} {\bibfnamefont {D.}~\bibnamefont
  {Calonico}}, \bibinfo {author} {\bibfnamefont {J.}~\bibnamefont {Catani}},
  \bibinfo {author} {\bibfnamefont {M.}~\bibnamefont {Inguscio}},\ and\
  \bibinfo {author} {\bibfnamefont {L.}~\bibnamefont {Fallani}},\ }\bibfield
  {title} {\bibinfo {title} {Synthetic dimensions and spin-orbit coupling with
  an optical clock transition},\ }\href@noop {} {\bibfield  {journal} {\bibinfo
   {journal} {Phys. Rev. Lett.}\ }\textbf {\bibinfo {volume} {117}},\ \bibinfo
  {pages} {220401} (\bibinfo {year} {2016})}\BibitemShut {NoStop}%
\bibitem [{\citenamefont {Riegger}\ \emph {et~al.}(2018)\citenamefont
  {Riegger}, \citenamefont {{Darkwah Oppong}}, \citenamefont {M}, \citenamefont
  {H\"{o}fer}, \citenamefont {Fernandes}, \citenamefont {Bloch},\ and\
  \citenamefont {F\"{o}lling}}]{LocalMagMom}%
  \BibitemOpen
  \bibfield  {author} {\bibinfo {author} {\bibfnamefont {L.}~\bibnamefont
  {Riegger}}, \bibinfo {author} {\bibfnamefont {N.}~\bibnamefont {{Darkwah
  Oppong}}}, \bibinfo {author} {\bibnamefont {M}}, \bibinfo {author}
  {\bibnamefont {H\"{o}fer}}, \bibinfo {author} {\bibfnamefont {D.~R.}\
  \bibnamefont {Fernandes}}, \bibinfo {author} {\bibfnamefont {I.}~\bibnamefont
  {Bloch}},\ and\ \bibinfo {author} {\bibfnamefont {S.}~\bibnamefont
  {F\"{o}lling}},\ }\bibfield  {title} {\bibinfo {title} {Localized magnetic
  moments with tunable spin exchange in a gas of ultracold fermions},\
  }\href@noop {} {\bibfield  {journal} {\bibinfo  {journal} {Phys. Rev. Lett.}\
  }\textbf {\bibinfo {volume} {120}},\ \bibinfo {pages} {143601} (\bibinfo
  {year} {2018})}\BibitemShut {NoStop}%
\bibitem [{\citenamefont {Scazza}\ \emph {et~al.}(2014)\citenamefont {Scazza},
  \citenamefont {Hofrichter}, \citenamefont {H\"{o}fer}, \citenamefont {Groot},
  \citenamefont {Bloch},\ and\ \citenamefont {F\"{o}lling}}]{ObsSpinEx}%
  \BibitemOpen
  \bibfield  {author} {\bibinfo {author} {\bibfnamefont {F.}~\bibnamefont
  {Scazza}}, \bibinfo {author} {\bibfnamefont {C.}~\bibnamefont {Hofrichter}},
  \bibinfo {author} {\bibfnamefont {M.}~\bibnamefont {H\"{o}fer}}, \bibinfo
  {author} {\bibfnamefont {P.~C.~D.}\ \bibnamefont {Groot}}, \bibinfo {author}
  {\bibfnamefont {I.}~\bibnamefont {Bloch}},\ and\ \bibinfo {author}
  {\bibfnamefont {S.}~\bibnamefont {F\"{o}lling}},\ }\bibfield  {title}
  {\bibinfo {title} {Observation of two-orbital spin-exchange interactions with
  ultracold {SU}(n)-symmetric fermions},\ }\href@noop {} {\bibfield  {journal}
  {\bibinfo  {journal} {Nature Physics}\ }\textbf {\bibinfo {volume} {10}},\
  \bibinfo {pages} {779–784} (\bibinfo {year} {2014})}\BibitemShut {NoStop}%
\bibitem [{\citenamefont {Wilde}(2017)}]{Wilde13}%
  \BibitemOpen
  \bibfield  {author} {\bibinfo {author} {\bibfnamefont {M.}~\bibnamefont
  {Wilde}},\ }\href@noop {} {\emph {\bibinfo {title} {Quantum Information
  Theory}}}\ (\bibinfo  {publisher} {Cambridge University Press},\ \bibinfo
  {year} {2017})\BibitemShut {NoStop}%
\bibitem [{\citenamefont {Neilsen}\ and\ \citenamefont
  {Cheung}(2012)}]{NielsenChueng12}%
  \BibitemOpen
  \bibfield  {author} {\bibinfo {author} {\bibfnamefont {M.~A.}\ \bibnamefont
  {Neilsen}}\ and\ \bibinfo {author} {\bibfnamefont {I.~L.}\ \bibnamefont
  {Cheung}},\ }\href@noop {} {\emph {\bibinfo {title} {Quantum Computation and
  Quantum Information}}}\ (\bibinfo  {publisher} {Cambridge University Press},\
  \bibinfo {year} {2012})\BibitemShut {NoStop}%
\bibitem [{\citenamefont {Navarrete-Benlloch}(2015)}]{QIcontvar}%
  \BibitemOpen
  \bibfield  {author} {\bibinfo {author} {\bibfnamefont {C.}~\bibnamefont
  {Navarrete-Benlloch}},\ }\href@noop {} {\emph {\bibinfo {title} {An
  Introduction to the Formalism of Quantum Information with Continuous
  Variables}}}\ (\bibinfo  {publisher} {Morgan \& Claypool Publishers},\
  \bibinfo {year} {2015})\BibitemShut {NoStop}%
\bibitem [{\citenamefont {Turner}\ \emph {et~al.}(2011)\citenamefont {Turner},
  \citenamefont {Pollmann},\ and\ \citenamefont {Berg}}]{TurnerPollmannBerg}%
  \BibitemOpen
  \bibfield  {author} {\bibinfo {author} {\bibfnamefont {A.~M.}\ \bibnamefont
  {Turner}}, \bibinfo {author} {\bibfnamefont {F.}~\bibnamefont {Pollmann}},\
  and\ \bibinfo {author} {\bibfnamefont {E.}~\bibnamefont {Berg}},\ }\bibfield
  {title} {\bibinfo {title} {Topological phases of one-dimensional fermions: An
  entanglement point of view},\ }\href@noop {} {\bibfield  {journal} {\bibinfo
  {journal} {Phys. Rev. B}\ }\textbf {\bibinfo {volume} {83}},\ \bibinfo
  {pages} {075102} (\bibinfo {year} {2011})}\BibitemShut {NoStop}%
\bibitem [{\citenamefont {Nakamura}(2012)}]{Nakamura}%
  \BibitemOpen
  \bibfield  {author} {\bibinfo {author} {\bibfnamefont {H.}~\bibnamefont
  {Nakamura}},\ }\href@noop {} {\emph {\bibinfo {title} {Nonadiabatic
  Transition: Concepts, Basic Theories, and Applications}}}\ (\bibinfo
  {publisher} {World Scientific},\ \bibinfo {year} {2012})\BibitemShut
  {NoStop}%
\bibitem [{\citenamefont {An}\ \emph {et~al.}(2017)\citenamefont {An},
  \citenamefont {Meier},\ and\ \citenamefont {Gadway}}]{DirectObsChiral}%
  \BibitemOpen
  \bibfield  {author} {\bibinfo {author} {\bibfnamefont {F.~A.}\ \bibnamefont
  {An}}, \bibinfo {author} {\bibfnamefont {E.~J.}\ \bibnamefont {Meier}},\ and\
  \bibinfo {author} {\bibfnamefont {B.}~\bibnamefont {Gadway}},\ }\bibfield
  {title} {\bibinfo {title} {Direct observation of chiral currents and magnetic
  reflection in atomic flux lattices},\ }\href@noop {} {\bibfield  {journal}
  {\bibinfo  {journal} {Science Advances}\ }\textbf {\bibinfo {volume} {3}},\
  \bibinfo {pages} {e1602685} (\bibinfo {year} {2017})}\BibitemShut {NoStop}%
\bibitem [{\citenamefont {Okunoa}\ \emph {et~al.}(2020)\citenamefont {Okunoa},
  \citenamefont {Amano}, \citenamefont {Enomoto}, \citenamefont {Takei1},\ and\
  \citenamefont {Takahashi}}]{SchemeSingleAtom}%
  \BibitemOpen
  \bibfield  {author} {\bibinfo {author} {\bibfnamefont {D.}~\bibnamefont
  {Okunoa}}, \bibinfo {author} {\bibfnamefont {Y.}~\bibnamefont {Amano}},
  \bibinfo {author} {\bibfnamefont {K.}~\bibnamefont {Enomoto}}, \bibinfo
  {author} {\bibfnamefont {N.}~\bibnamefont {Takei1}},\ and\ \bibinfo {author}
  {\bibfnamefont {Y.}~\bibnamefont {Takahashi}},\ }\bibfield  {title} {\bibinfo
  {title} {Schemes for nondestructive quantum gas microscopy of single atoms in
  an optical lattice},\ }\href@noop {} {\bibfield  {journal} {\bibinfo
  {journal} {New J. Phys.}\ }\textbf {\bibinfo {volume} {22}},\ \bibinfo
  {pages} {013041} (\bibinfo {year} {2020})}\BibitemShut {NoStop}%
\bibitem [{\citenamefont {Takata}\ \emph {et~al.}(2019)\citenamefont {Takata},
  \citenamefont {Nakajima}, \citenamefont {Kobayashi}, \citenamefont {Ono},
  \citenamefont {Amano},\ and\ \citenamefont {Takahashi}}]{YbClockTrans}%
  \BibitemOpen
  \bibfield  {author} {\bibinfo {author} {\bibfnamefont {Y.}~\bibnamefont
  {Takata}}, \bibinfo {author} {\bibfnamefont {S.}~\bibnamefont {Nakajima}},
  \bibinfo {author} {\bibfnamefont {J.}~\bibnamefont {Kobayashi}}, \bibinfo
  {author} {\bibfnamefont {K.}~\bibnamefont {Ono}}, \bibinfo {author}
  {\bibfnamefont {Y.}~\bibnamefont {Amano}},\ and\ \bibinfo {author}
  {\bibfnamefont {Y.}~\bibnamefont {Takahashi}},\ }\bibfield  {title} {\bibinfo
  {title} {Current-feedback-stabilized laser system for quantum simulation
  experiments using {Y}b clock transition at 578 nm},\ }\href@noop {}
  {\bibfield  {journal} {\bibinfo  {journal} {Rev. Sci. Instrum.}\ }\textbf
  {\bibinfo {volume} {90}},\ \bibinfo {pages} {083002} (\bibinfo {year}
  {2019})}\BibitemShut {NoStop}%
\bibitem [{\citenamefont {Fendley}(2016)}]{Fendley}%
  \BibitemOpen
  \bibfield  {author} {\bibinfo {author} {\bibfnamefont {P.}~\bibnamefont
  {Fendley}},\ }\bibfield  {title} {\bibinfo {title} {Strong zero modes and
  eigenstate phase transitions in the {XYZ}/interacting {M}ajorana chain},\
  }\href@noop {} {\bibfield  {journal} {\bibinfo  {journal} {J. Phys. A: Math.
  Theor.}\ }\textbf {\bibinfo {volume} {49}},\ \bibinfo {pages} {30LT01}
  (\bibinfo {year} {2016})}\BibitemShut {NoStop}%
\bibitem [{\citenamefont {Maceira}\ and\ \citenamefont
  {Mila}(2018)}]{Maceira_2018}%
  \BibitemOpen
  \bibfield  {author} {\bibinfo {author} {\bibfnamefont {I.~A.}\ \bibnamefont
  {Maceira}}\ and\ \bibinfo {author} {\bibfnamefont {F.}~\bibnamefont {Mila}},\
  }\bibfield  {title} {\bibinfo {title} {Infinite coherence time of edge spins
  in finite-length chains},\ }\href
  {https://doi.org/10.1103/PhysRevB.97.064424} {\bibfield  {journal} {\bibinfo
  {journal} {Phys. Rev. B}\ }\textbf {\bibinfo {volume} {97}},\ \bibinfo
  {pages} {064424} (\bibinfo {year} {2018})}\BibitemShut {NoStop}%
\bibitem [{\citenamefont {Kemp}\ \emph {et~al.}(2017)\citenamefont {Kemp},
  \citenamefont {Yao}, \citenamefont {Laumann},\ and\ \citenamefont
  {Fendley}}]{Kemp_2017}%
  \BibitemOpen
  \bibfield  {author} {\bibinfo {author} {\bibfnamefont {J.}~\bibnamefont
  {Kemp}}, \bibinfo {author} {\bibfnamefont {N.~Y.}\ \bibnamefont {Yao}},
  \bibinfo {author} {\bibfnamefont {C.~R.}\ \bibnamefont {Laumann}},\ and\
  \bibinfo {author} {\bibfnamefont {P.}~\bibnamefont {Fendley}},\ }\bibfield
  {title} {\bibinfo {title} {Long coherence times for edge spins},\ }\href
  {https://doi.org/10.1088/1742-5468/aa73f0} {\bibfield  {journal} {\bibinfo
  {journal} {Journal of Statistical Mechanics: Theory and Experiment}\ }\textbf
  {\bibinfo {volume} {2017}},\ \bibinfo {pages} {063105} (\bibinfo {year}
  {2017})}\BibitemShut {NoStop}%
\bibitem [{\citenamefont {Monthus}(2018)}]{Monthus_2018}%
  \BibitemOpen
  \bibfield  {author} {\bibinfo {author} {\bibfnamefont {C.}~\bibnamefont
  {Monthus}},\ }\bibfield  {title} {\bibinfo {title} {Even and odd normalized
  zero modes in random interacting {M}ajorana models respecting the parity {P}
  and the time-reversal-symmetry {T}},\ }\href
  {https://doi.org/10.1088/1751-8121/aac4b0} {\bibfield  {journal} {\bibinfo
  {journal} {Journal of Physics A: Mathematical and Theoretical}\ }\textbf
  {\bibinfo {volume} {51}},\ \bibinfo {pages} {265303} (\bibinfo {year}
  {2018})}\BibitemShut {NoStop}%
\end{thebibliography}%
\providecommand{\noopsort}[1]{}\providecommand{\singleletter}[1]{#1}%

\end{document}